\documentclass[twocolumn,superscriptaddress,showkeys,aps,pop,floatfix]{revtex4-2}
\usepackage{amsmath}
\usepackage{amssymb}
\usepackage{upgreek}
\usepackage{graphicx}
\usepackage{adjustbox}
\usepackage{float}
\usepackage{rotating}
\usepackage{multirow}
\usepackage{booktabs}
\usepackage{dcolumn}
\usepackage{xcolor}
\usepackage{color}
\usepackage[normalem]{ulem}

\usepackage{hyperref}
\hypersetup{
    colorlinks=true,
    citecolor=blue,
    linkcolor=blue,
    filecolor=blue,
    urlcolor=blue,
}

\usepackage{natbib}
\newcommand{\Smilei}{{\sc Smilei }}

\begin{document}

\title{Electron injection and acceleration into laser-driven wakefield from a solid overdense plasma target}

\def\LSI{LSI, CEA/DRF/IRAMIS, CNRS, \'Ecole Polytechnique, Institut Polytechnique de Paris, F-91120 Palaiseau, France}

\author{M. Caetano de Sousa}
\thanks{Current affiliation: Centre Lasers Intenses et Applications (CELIA), Universit\'e de Bordeaux-CNRS-CEA, UMR 5107, F-33405 Talence, France}
\affiliation\LSI

\author{S. Marini}
\affiliation{CEA, IRFU, DACM, Université Paris-Saclay, 91191 Gif-sur-Yvette, France}

\author{M. Grech}
\affiliation{LULI, CNRS, CEA, Sorbonne Universit\'e, \'Ecole Polytechnique, Institut Polytechnique de Paris, F-91120 Palaiseau, France}

\author{S. Brunner}
\affiliation{Swiss Plasma Center, \'Ecole Polytechnique F\'ed\'erale de Lausanne (EPFL), CH-1015 Lausanne, Switzerland}

\author{C. Riconda}
\affiliation{LULI, Sorbonne Universit\'e, CNRS, CEA, \'Ecole Polytechnique, Institut Polytechnique de Paris, F-75252 Paris, France}

\author{M. Raynaud}
\affiliation\LSI

\begin{abstract}
A laser-plasma acceleration scheme combining electron extraction from a solid overdense target with wakefield acceleration in an adjacent underdense plasma region is presented. A laser pulse excites a diffracted electromagnetic wave at the overdense plasma interface, extracting and pre-accelerating electrons, which are then injected into laser-driven wakefield cavities in the underdense plasma. A parametric study identifies key conditions enabling efficient electron injection and energy gain in this two stage acceleration configuration. Two-dimensional particle-in-cell simulations performed with the \Smilei code show that the proposed scheme produces electron bunches with a tunable trade-off between charge, energy, and beam quality at laser intensity $I_0 \lambda_0^2 \simeq 3.4 \times 10^{19}$~W$\mu$m$^2$/cm$^2$ ($\lambda_0=0.8~\mu$m). According to the parameters used, the electron beam is accelerated to peak energies of $\sim150-250$~MeV with an estimated charge in 3D of $\sim50-400$~pC integrated over the full width at half maximum energy range, and $\sim100-1900$~pC with energies above $50$~MeV.
\end{abstract}

\maketitle
\date{\today}


\section{Introduction}
\label{Sec:Introduction}

Laser–plasma acceleration has emerged as a promising technique to generate high-energy electron beams (up to $\sim$GeV) over short distances ($\sim$mm) in both underdense~\cite{Tajima1979, Esarey2009, McGuffey2010, Pak2010, Chen2012, Couperus2017, Lee2018, Gonsalves2019, Kirchen2021, Jalas2023, Winkler2025} and overdense plasmas~\cite{Wilks1992, Mora2003, Naumova2004, Kupersztych2004, Naumova2005, Raynaud2007, Arefiev2015, Riconda2015, Thevenet2016, Fedeli2016, Macchi2018, Cantono2018, Gong2019, Wen2020, Jirka2020, Raynaud2020, Marini2021A, Marini2021B, Shen2021B, Sarma2022, Singh2022, Marini2023}, defined respectively by the plasma density $n_e$ below and above the critical density $n_c$ for a laser pulse of frequency $\omega_0$. Each regime leads to electron beams with distinct characteristics.

Laser interaction with overdense plasmas can result in electron acceleration through various mechanisms, including vacuum laser acceleration (VLA)~\cite{Naumova2004, Naumova2005, Singh2022}, direct laser acceleration (DLA)~\cite{Arefiev2015, Thevenet2016, Gong2019, Wen2020, Jirka2020}, as well as resonantly excited relativistic surface plasma waves (SPWs)~\cite{Kupersztych2004, Raynaud2007, Riconda2015, Fedeli2016, Macchi2018, Cantono2018, Raynaud2020, Marini2021A, Marini2021B, Shen2021B}. Notably, when the laser interacts with the plasma edge, both SPWs~\cite{Shen2021A, Shen2021B, Sarma2022} and diffracted waves~\cite{Marini2023} are excited, each contributing differently to the electron acceleration process.

Diffracted waves, specifically, generate strong longitudinal electric fields that propagate in vacuum at the speed of light and decay with the longitudinal distance. These fields enable efficient electron acceleration, with energy gain $\gamma$ scaling as $\gamma=2 \eta a_0\sqrt{k_0 \Delta x}$, where $a_0$ is the normalized laser amplitude, $k_0$ is the wave number, $\Delta x$ is the electron displacement from the interaction point, and $\eta \lesssim 1$ (obtained from simulations) is the ratio between the maximum amplitudes of diffracted and laser fields~\cite{Marini2023}.

Laser-overdense plasma interaction mechanisms can drive electron beams with substantial charge (up to nC) and energies in the MeV range over short distances, with acceleration gradients reaching the TeV/m level. However, the resulting electron spectra are typically continuous rather than monoenergetic, limiting their use in applications requiring well-defined beams. In contrast, laser interaction with underdense plasmas excites plasma wakefields that can capture and accelerate electrons up to GeV energies with small energy spread and pC of charge~\cite{Gonsalves2019, Winkler2025}.

Efficient acceleration in underdense plasmas depends on appropriate electron injection into the accelerating phase of the wake. Wave breaking, for example, can lead to electron self-injection~\cite{Bulanov1998} through a suitable plasma density profile. Injection can also be independently controlled by means of colliding laser pulses~\cite{Esarey1997PRL, Faure2006} or ionization in doped plasmas~\cite{RowlandsRees2008, McGuffey2010, Pak2010, Chen2012}.

To improve injection into wakefields, alternative configurations involving multi-region targets have also been investigated. Zhang~\textit{et al.}~\cite{Zhang2015} proposed placing a thin ($\sim\mu$m) near-critical density layer ($n_e = 4n_c$) in front of an underdense plasma. In this arrangement, the laser pulse extracts electrons from the layer, which are subsequently injected and accelerated in the wakefield excited in the underdense region. Fedeli \textit{et al.}~\cite{Fedeli2022} proposed obliquely irradiating an overdense plasma-mirror surface with a laser pulse to extract electrons, which are subsequently accelerated in a nearby underdense plasma by laser-driven wakefield.

This work investigates an integrated laser-plasma acceleration arrangement that combines laser-overdense and laser-underdense plasma interactions. A single laser pulse excites both a diffracted wave at the overdense plasma interface and a wakefield in the adjacent underdense region. The laser-excited diffracted waves efficiently extract and pre-accelerate electron bunches from the overdense plasma, which are subsequently injected into the laser-driven wakefield for further acceleration. Electrons from the underdense plasma are also self-injected and accelerated by the wakefield. Since electrons originating from both the overdense and the underdense plasma interact with each other, their injection and self-injection processes are not completely independent. Moreover, the proposed scheme uses a single laser pulse to extract electrons from the overdense target and to drive the wakefield in the underdense region, such that the electron extraction, injection and acceleration stages are intrinsically coupled.

Two-dimensional (2D3V) particle-in-cell simulations were carried out for different parameters values. They demonstrate the generation of high-quality, high-energy electron bunches reaching $\sim150-250$~MeV at peak over a propagation distance of $\sim685-975 \lambda_0$ (where $\lambda_0$ is the laser wavelength), with three-dimensional charges, estimated by a geometric extrapolation of the 2D results, of $\sim50-400$~pC in the full width at half maximum (FWHM) energy range, $\sim100-1900$~pC with energies above $50$~MeV, and $\sim100-1250$~pC with energies above $100$~MeV.

This paper is organized as follows: Section~\ref{Sec:Setup} introduces the setup for the proposed acceleration mechanism and presents the simulations design. Section~\ref{Sec:InjectionAcceleration} analyzes the injection and acceleration processes, as well as the electron dynamics. Section~\ref{Sec:ParametricStudy} presents a parametric study identifying the optimum conditions for the proposed acceleration mechanism. Section~\ref{Sec:OptimizedAcceleration} details the optimized configurations for electron beam acceleration aiming either at higher energy or higher charge. Finally, Section~\ref{Sec:Conclusion} summarizes the findings and discusses possibilities for future works.


\section{Setup and Particle-in-Cell Simulations}
\label{Sec:Setup}

\begin{figure}[tb]
    \centering
    \includegraphics[width=0.9\linewidth]{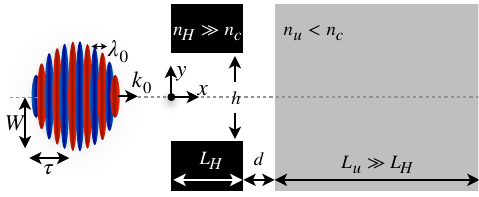}
    \caption{Laser-plasma interaction setup.}
    \label{fig:LaserPlasmaSetup}
\end{figure}

The proposed setup is depicted in Fig.~\ref{fig:LaserPlasmaSetup}. A $p$-polarized ({\it i.e.} electric field parallel to the plane of incidence) Gaussian laser pulse propagates in the $x$-direction with wavelength $\lambda_0$, angular frequency $\omega_0$ and period $\tau_0=\lambda_0/c$, where $c$ is the speed of light in vacuum and we consider $\lambda_0 = 0.8~\mu$m. The laser has a waist (minimum radius) $W = 9\lambda_0 = 7.2~\mu$m, full width at half maximum (FWHM) pulse duration $\tau = 5\tau_0\approx 13.3$~fs, and normalized vector potential $a_0 = 5$, with $a_0 \equiv e E_0/(m_e c \omega_0)$, $E_0$ the peak electric field amplitude of the laser pulse, $e$ the electron charge and $m_e$ the electron mass. These parameters correspond to a laser intensity $I_0 \lambda_0^2 \simeq 3.4 \times 10^{19}$~W$\mu$m$^2$/cm$^2$. 

The laser sequentially interacts with two plasma regions of different densities. It first propagates through the cylindrical aperture of an overdense plasma target with length $L_H$ ($0 < x \leq L_H$), aperture diameter $h$ in the $y$O$z$ plane (transverse direction), density $n_H = 100n_c \simeq 1.7 \times 10^{23}$~cm$^{-3}$, and plasma wavelength $\lambda_H \simeq \lambda_0 / \sqrt{n_H/n_c} = 80$~nm. Here $n_c = \varepsilon_0 m_e \omega_0^2 / e^2$ is the critical plasma density, and $\varepsilon_0$ is the vacuum permittivity. The laser pulse then crosses a vacuum region (gap) of length $d$ ($L_H < x \leq L_H + d$) before entering a uniform underdense plasma with density $n_u \ll n_c$ extending from $x = L_H + d$ to $x = L_H + d + L_u$. The laser is focused at the center of the underdense plasma region $x_f = L_H + d + L_u/2$, enabling a two-stage acceleration mechanism.

To investigate the injection and acceleration dynamics, two-dimensional (2D3V) particle-in-cell (PIC) simulations were conducted using the open-source code \Smilei~\cite{Derouillat2018}. In the simulations, both plasma regions consist of cold electrons with $T_e \to 0$ and a neutralizing background of cold ions with atomic number $Z = 1$. The simulation box extends over $334\lambda_0$ ($=267~\mu$m) in the $x$-direction and $192\lambda_0$ ($=154~\mu$m) in the $y$-direction. The spatial resolution is set to $\Delta x = \Delta y = \lambda_0/128 = \lambda_H / 12.8$ ($=6.25$~nm), such that it is enough to resolve the overdense plasma wave with more than ten points. The simulation time step is chosen to be $c \Delta t = 0.95 \Delta x/\sqrt{2}$, corresponding to $95$\% of the Courant–Friedrich–Lewy (CFL) condition for the standard finite-difference time-domain (FDTD) solver~\cite{Nuter2014}. Each cell initially contains $124$ randomly distributed numerical particles of each species in the overdense plasma region and $4$ numerical particles of each species in the underdense plasma. Particle boundary conditions in $x$ are removing (left) or thermalizing (right), and thermalizing in the $y$-direction. It means that particles reaching the left-$x$ boundary are deleted, while particles reaching all the other boundaries have their temperature set to $T \to 0$. Electromagnetic field boundary conditions are perfectly matched layers (PML), which are open boundary in $y$ and allow laser injection (absorption) from the left (right) $x$ boundary. The final simulation time was set to $t_f = 349\uptau_0 \approx 931~$fs.


\begin{figure*}[tb]
\centering
   \includegraphics[width=1.0\linewidth]{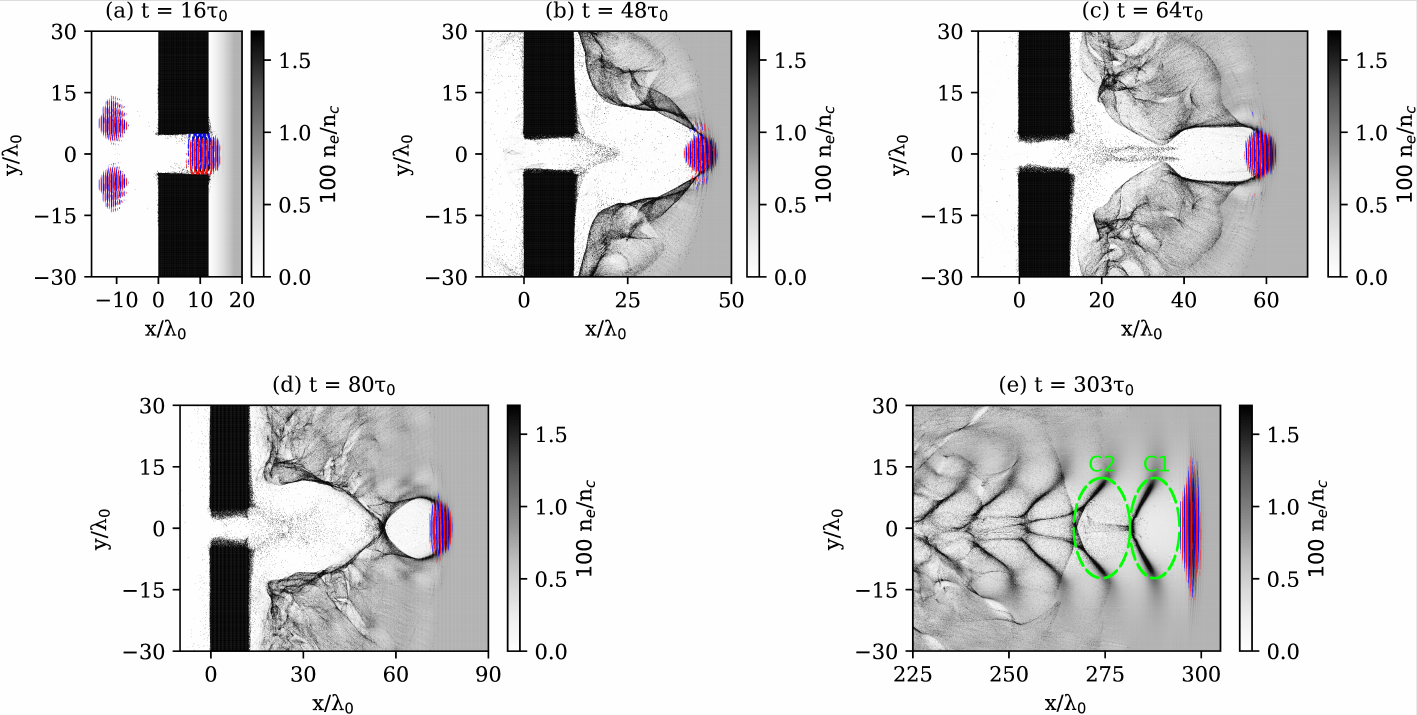}
    \caption{Spatiotemporal evolution of laser-plasma interaction showing the electron density (grayscale) and the magnetic field component $B_z$ of the laser pulse (red/blue) in $x$-$y$ space at different instants of time: (a)~part of the laser pulse crosses and exits the overdense target aperture and part of the laser is reflected by the overdense target around its aperture at $t=16\tau_0$, (b)~laser entering the underdense plasma region at $t=48\tau_0$, (c)~formation of the first wakefield cavity and electron capture at $t=64\tau_0$, (d)~closure of the first wakefield cavity at $t=80\tau_0$, and (e)~multiple wakefield cavities at $t=303\tau_0$ with the first (C1) and second (C2) cavities indicated by the dashed ellipses in green.}
\label{fig:ElectronDensityEvolution}
\end{figure*}

\section{Injection and Acceleration Mechanism}
\label{Sec:InjectionAcceleration}

\begin{figure}[tb]
\centering
    \includegraphics[width=0.74\linewidth]{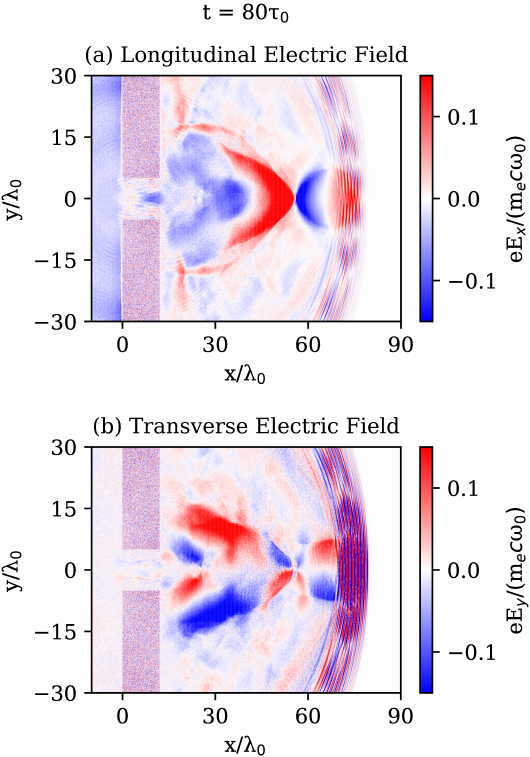}
    \caption{Electric field in the (a) longitudinal $x-$direction and in the (b) transverse $y-$direction at $t=80\tau_0$.}
    \label{fig:ElectricField80t0}
\end{figure}

\begin{figure}[tb]
\centering
    \includegraphics[width=0.74\linewidth]{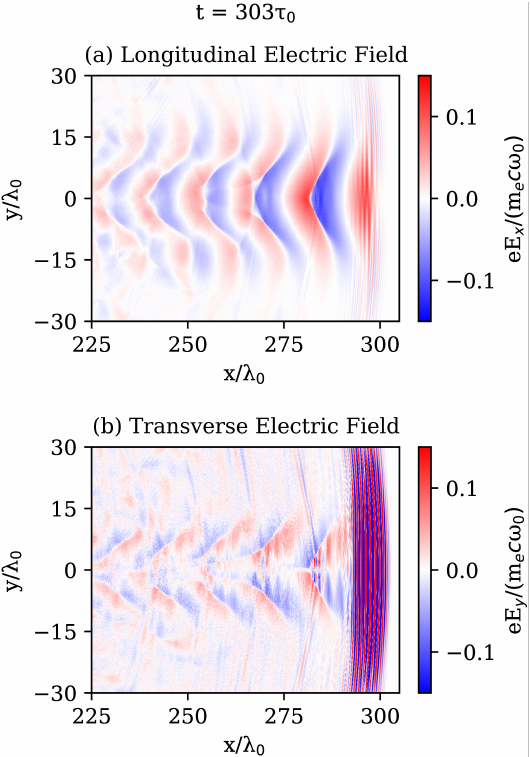}
    \caption{Electric field in the (a) longitudinal $x-$direction and in the (b) transverse $y-$direction at $t=303\tau_0$.}
    \label{fig:ElectricField303t0}
\end{figure}

The electron dynamics resulting from the laser-plasma interaction setup discussed in Section~\ref{Sec:Setup} is here investigated assuming $L_H = 12 \lambda_0 = 9.6~\mu$m, $h = 10 \lambda_0 = 8~\mu$m, $d = 0$, and $n_u = 0.007n_c \simeq 1.2 \times 10^{19}$~cm$^{-3}$. At $t=0$, the laser pulse reaches the overdense plasma target edge located at $x = 0$. It is focused at $x_f \gg L_H$, such that the laser beam waist in the overdense target region exceeds the aperture radius $h/2$. This geometry leads to a direct interaction of the laser pulse with the overdense plasma edge, exciting diffracted electromagnetic waves that efficiently extract and accelerate electrons from the plasma surface, as depicted in Fig.~\ref{fig:ElectronDensityEvolution}(a) at $t=16 \tau_0$. From Ref.~\cite{Marini2023}, electrons accelerated by such diffracted waves achieve energies up to $\gamma = 2 \eta  a_0 \sqrt{k_0 L_H}$, with $\eta \lesssim 1$ the ratio between the maximum amplitudes of the diffracted waves and the laser electromagnetic field. In our simulations, electron energies of approximately $28$~MeV ($\gamma \approx 55$) are observed, which is slightly weaker than in Ref~\cite{Marini2023} where the laser is focused at the plasma edge such that the interaction happens at the laser maximum intensity.

Besides being accelerated, electrons from the overdense plasma are also heated by the laser, and remain co-propagating with it as the pulse advances beyond the target, as can be seen in Figs.~\ref{fig:ElectronDensityEvolution}(b)-(d). At $t=48\tau_0$, Fig.~\ref{fig:ElectronDensityEvolution}(b), the laser pulse has entered the underdense plasma region, with pre-accelerated and pre-heated electrons from the overdense plasma propagating behind it. As the laser advances through the underdense plasma, it exerts a ponderomotive force on the background electrons, forming characteristic wakefield cavities, as illustrated in Figs.~\ref{fig:ElectronDensityEvolution}(c)-(d) at $t=64\tau_0$ and $t=80\tau_0$. Persistent electromagnetic fields continue to extract electrons from the overdense target, sustaining their injection into the wakefield structure. By $t=303\tau_0$, Fig.~\ref{fig:ElectronDensityEvolution}(e) shows that multiple well-defined wakefield cavities are present, all of which can support the formation and acceleration of distinct electron bunches.

Figures~\ref{fig:ElectricField80t0} and \ref{fig:ElectricField303t0} illustrate the electric field structure of the wakefield at two instants of time: $t=80\tau_0$ and $t=303\tau_0$ respectively. The longitudinal field $E_x$ is shown in the upper panels Figs.~\ref{fig:ElectricField80t0}(a) and \ref{fig:ElectricField303t0}(a), while the transverse field $E_y$ is presented in the lower panels Figs.~\ref{fig:ElectricField80t0}(b) and \ref{fig:ElectricField303t0}(b). At $t=80\tau_0$, strong longitudinal fields create an accelerating gradient that captures and accelerates electrons coming from the overdense target into the dark blue regions of Fig.~\ref{fig:ElectricField80t0}(a). Simultaneously, the transverse fields provide the focusing forces necessary for beam confinement, as illustrated in Fig.~\ref{fig:ElectricField80t0}(b). By $t=303\tau_0$, the electric field structures stabilize throughout multiple wakefield periods, maintaining both the acceleration mechanism and radial focusing required for electron bunch formation, as shown in Figs.~\ref{fig:ElectricField303t0}(a) and \ref{fig:ElectricField303t0}(b).

\begin{figure}[bt]
    \centering
    \includegraphics[width=1.0\linewidth]{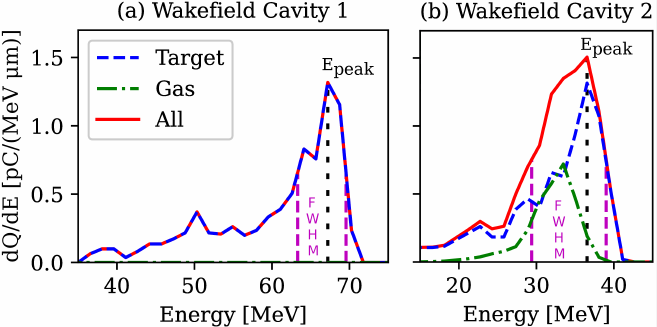}
    \caption{Electron energy spectra in the (a) first and (b) second wakefield cavities at $t = 303\uptau_0$. The dashed blue curve represents electrons from the overdense plasma (target), while the dash-dotted green curve shows electrons from the underdense plasma (gas). The solid red curve indicates the total electron population (all). 
    The vertical dotted black line marks the peak energy, and the vertical dashed violet lines indicate the FWHM energy range.}
    \label{fig:ChargeDistribution303t0}
\end{figure}

At $t = 303\tau_0$, most of the charge is concentrated in the two first cavities defined as C1 and C2 in Fig.~\ref{fig:ElectronDensityEvolution}(e). Thus, in the following, the discussion will concentrate on the first and second wakefield cavities. Their charge distribution as a function of the electron energy at $t = 303\tau_0$ is shown in Fig.~\ref{fig:ChargeDistribution303t0}. There, the solid red line represents the total electron population (all), the dashed blue line shows electrons originating from the overdense plasma (target), and the dash-dotted green line indicates electrons from the underdense plasma (gas). In the first cavity (C1), Fig.~\ref{fig:ChargeDistribution303t0}(a), all accelerated electrons originate from the overdense plasma, whereas in the second cavity (C2), Fig.~\ref{fig:ChargeDistribution303t0}(b), a combination of electrons from both overdense and underdense plasma is observed. It demonstrates that, in the present configuration, the injection and acceleration of both electron populations in the wakefield cavities are not independent, but rather intrinsically coupled. Furthermore, the acceleration of electron bunches with a significant amount of particles from the overdense target validates the proposed setup as an alternative and efficient method for self-consistent electron injection into laser-driven wakefields.

Both cavities present a clean spectrum with a well defined peak in energy: $67$~MeV with a FWHM spread of $6.2$~MeV in the first cavity, and $37$~MeV with a FWHM spread of $9.6$~MeV in the second one. Nonetheless, the beam quality decreases in successive wakefield cavities, consistent with the expected degradation of the wakefield structure. Here, the beam quality is analyzed in terms of its FWHM energy spread, transverse size in the $y$ direction, divergence (angle with respect to the $x$ direction) and normalized rms (root mean square) emittance given by
\begin{equation}
    \text{emittance}_{\text{norm}} = \sqrt{\sigma^2(y) \sigma^2(p_y) - \langle y \, p_y \rangle ^2} \, ,
\end{equation}
where $\sigma^2(y)$, $\sigma^2(p_y)$ are the variances of $y$ and $p_y$, and $\langle y \, p_y \rangle$ is the expected value of $y \, p_y$.

In the first cavity, the bunch reaches its highest peak energy ($67$~MeV) with a minimal relative spread ($9.3\%$), forming a compact and spatially homogeneous beam (see Table~\ref{tab:beam_params_303t0}). All accelerated electrons originate from the overdense target. It yields lower FWHM divergence ($168$~mrad) and FWHM transverse size ($5.1$~$\mu$m), and a moderate FWHM normalized emittance ($5.6$~$\pi$~mm~mrad). In the second cavity, the wake potential weakens: the peak energy is smaller ($37$~MeV) and the relative energy spectrum is broader, with a FWHM energy spread of $26\%$. The lower peak energy and the larger FWHM divergence ($253$~mrad) and FWHM transverse size ($7.7$~$\mu$m) reflect the partial wake degradation in the second cavity, but the bunch remains homogeneous and the normalized emittance is smaller ($4.0$~$\pi$~mm~mrad).

We calculate the charge in the wakefield cavities in two ways: either by considering all the charge having an energy greater than 30~MeV (which is more or less half of the peak energy in the first cavity) as in Ref.~\cite{Marini2023}, or by integrating the charge over the FWHM energy range shown in Fig.~\ref{fig:ChargeDistribution303t0}. The FWHM charge accounts for the spectrum useful for applications, such as medical, requiring low energy dispersion of the beam. While the charge above a certain energy level, 30~MeV for example, is important for applications requiring higher amounts of charge irrespective of the beam energy dispersion.

The total charge in 3D is next evaluated for both cases as a geometric extrapolation of the 2D results considering that electrons from the overdense plasma are emitted from a cylindrical aperture of radius $r_H = h/2$. The cylindrical aperture emitting electrons from the overdense plasma has a circumference equal to $2 \pi r_H$ in the $y$O$z$ plane. While the underdense wakefield cavity has a transverse section of area $\pi r_u^2$ in the $y$O$z$ plane, with $r_u$ the wakefield cavity radius in $y$ obtained in the simulations. Thus, the charge in 3D is estimated by multiplying i) the 2D charge in pC/$\mu$m (integrated from the simulations in the $x$O$y$ plane) originating from the overdense plasma by $\pi r_H$ and ii) the 2D charge in pC/$\mu$m coming from the underdense plasma by $\pi r_u/2$. It is important to note that the 3D charge estimate considers a semi-circumference and a semi-circle area because the charge in 2D obtained from the simulations takes into account the overdense plasma aperture diameter $h$, as well as the underdense plasma wakefield cavity diameter ($= 2r_u$) in the $y$ direction.

\setlength{\tabcolsep}{8.0pt}
\renewcommand{\arraystretch}{1.15}
\begin{table}[tb]
\caption{Beam parameters in the first (C1) and second (C2) wakefield cavities at $t = 303\tau_0$.}
  \centering
  \footnotesize
  \begin{tabular}{lcc}
    \toprule
    \textbf{Parameter} & \textbf{C1} & \textbf{C2} \\
    \midrule
    Maximum Energy (MeV)                                & 71    & 42  \\
    Peak Energy (MeV)                                   & 67    & 37  \\
    FWHM Energy Spread (MeV)                            & 6.2   & 9.6 \\
    FWHM Energy Spread ($\%$)                           & 9.3   & 26  \\
    FWHM Divergence (mrad)                              & 168   & 253 \\
    FWHM Transverse Size in $y$ direction ($\mu$m)      & 5.1   & 7.7 \\
    FWHM Normalized Emittance ($\pi$ mm mrad)           & 5.6   & 4.0 \\
    Integrated FWHM Charge (pC/$\mu$m)                  & 4.1   & 7.7 \\
    \hspace{10.0pt} From Overdense Plasma ($\%$)        & 100   & 67  \\
    \hspace{10.0pt} From Underdense Plasma ($\%$)       & 0     & 33  \\
    Estimated Total FWHM Charge (pC)                    & 52    & 98  \\
    Integrated Charge above 30~MeV (pC/$\mu$m)          & 8.1   & 7.8 \\
    \hspace{10.0pt} From Overdense Plasma ($\%$)        & 100   & 70  \\
    \hspace{10.0pt} From Underdense Plasma ($\%$)       & 0     & 30  \\
    Estimated Total Charge above 30~MeV (pC)            & 101   & 100 \\
    \bottomrule
  \end{tabular}
  \label{tab:beam_params_303t0}
\end{table}

The charge values are reported in Table~\ref{tab:beam_params_303t0}. The first wakefield cavity accelerates a charge of $4.1$~pC/$\mu$m in the FWHM energy range, and $8.1$~pC/$\mu$m with energies above 30~MeV. The FWHM charge accelerated in the second cavity is higher since now the electrons come from both the overdense ($\sim 67\%$) and the underdense ($\sim 33\%$) plasma: $7.7$~pC/$\mu$m. On the other hand, the charge with energies above 30~MeV is lower ($7.8$~pC/$\mu$m) in the second cavity since the peak and maximum energies are also lower due to the partial wake degradation in this cavity.


\begin{figure*}[bt]
    \centering
    \includegraphics[width=0.95\linewidth]{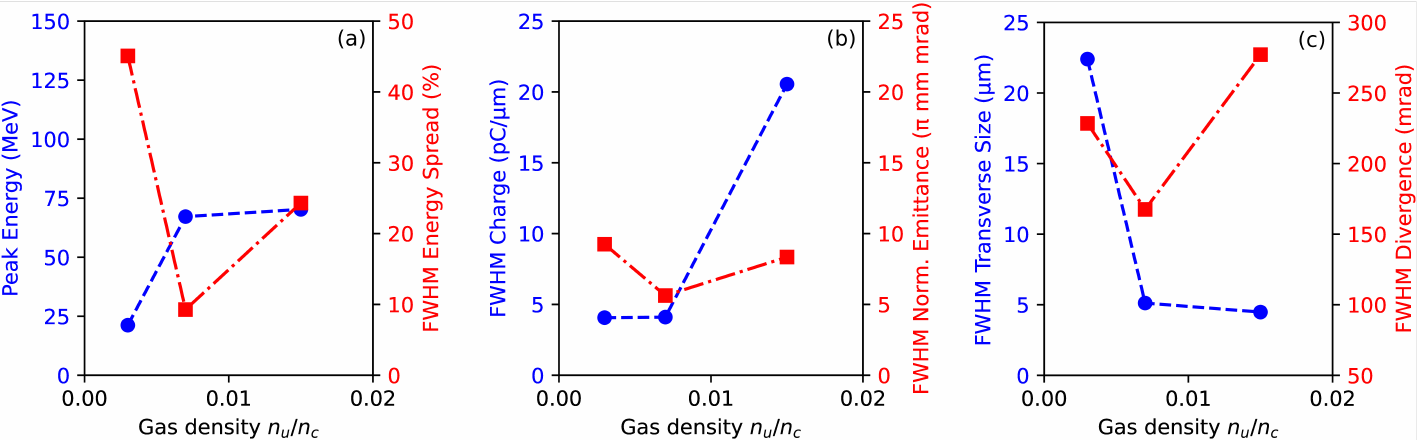}
    \caption{Parametric study as a function of the normalized underdense plasma density $n_u/n_c$ for electrons accelerated in the first wakefield cavity: (a)~bunch peak energy (blue circles) and FWHM energy spread (red squares), (b)~FWHM beam charge (blue circles) and FWHM normalized emittance (red squares), and (c)~FWHM transverse size (blue circles) and FWHM divergence (red squares). The fixed parameters are $d = 0$, $L_H = 12 \lambda_0 = 9.6~\mu$m and $h = 10 \lambda_0 = 8~\mu$m. The laser pulse has propagated a distance of $\sim24\lambda_u$ in the underdense plasma region, which corresponds to a propagation time of $t = 457\tau_0$ for $n_u / n_c = 0.003$, $t = 303\tau_0$ for $n_u / n_c = 0.007$, and $t = 214\tau_0$ for $n_u / n_c = 0.015$.}
\label{fig:ParametricStudy_nu}
\end{figure*}

\section{Parametric Study of the Laser-Plasma Interaction}
\label{Sec:ParametricStudy}

\begin{figure*}[bt]
    \centering
    \includegraphics[width=0.95\linewidth]{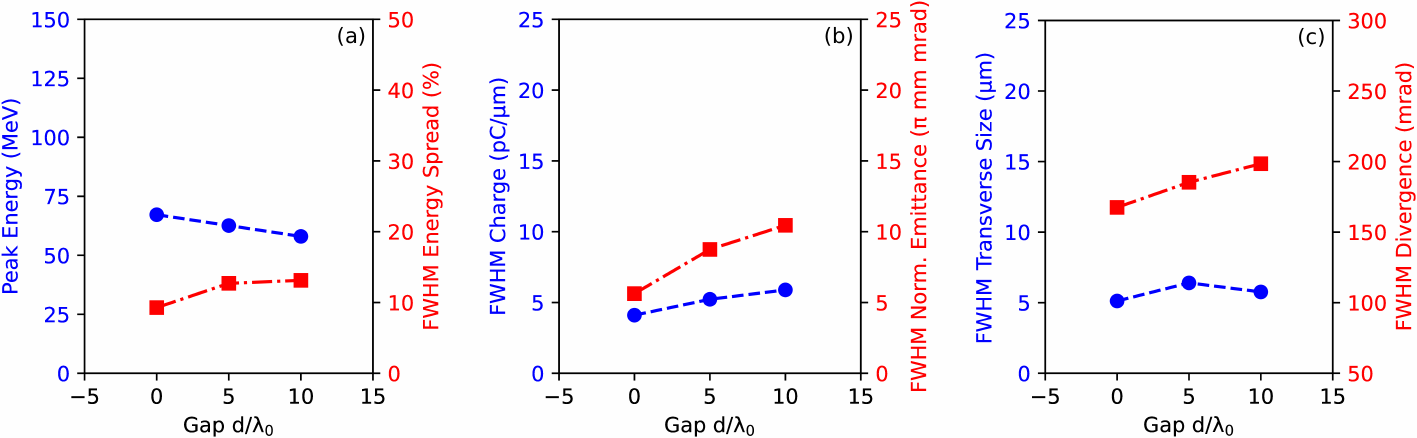}
    \caption{Parametric study as a function of the normalized gap distance $d / \lambda_0$ between the overdense and underdense plasma for electrons accelerated in the first wakefield cavity: (a)~bunch peak energy (blue circles) and FWHM energy spread (red squares), (b)~FWHM beam charge (blue circles) and FWHM normalized emittance (red squares), and (c)~FWHM transverse size (blue circles) and FWHM divergence (red squares). The fixed parameters are $n_u = 0.007n_c \simeq 1.2 \times 10^{19}$~cm$^{-3}$, $L_H = 12 \lambda_0 = 9.6~\mu$m and $h = 10 \lambda_0 = 8~\mu$m. The laser pulse has propagated a distance of $\sim24\lambda_u$ in the underdense plasma region, which corresponds to a total propagation time of $t = 303\tau_0$ for $d = 0$, $t = 308\tau_0$ for $d = 5 \lambda_0$, and $t = 313\tau_0$ for $d = 10 \lambda_0$.}
\label{fig:ParametricStudy_d}
\end{figure*}

\begin{figure*}[bt]
    \centering
    \includegraphics[width=0.95\linewidth]{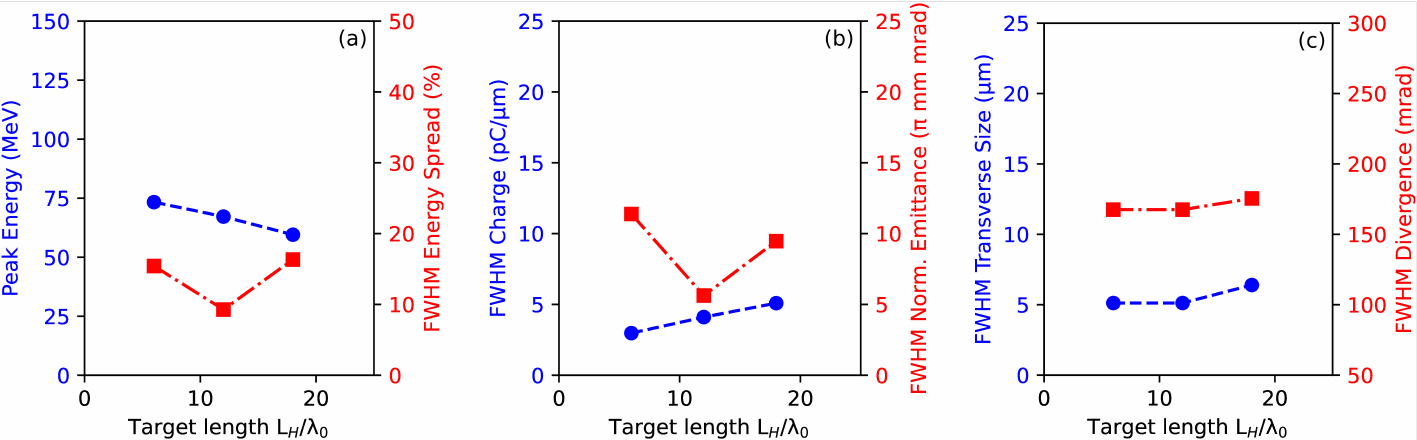}
    \caption{Parametric study as a function of the normalized overdense plasma length $L_H / \lambda_0$ for electrons accelerated in the first wakefield cavity: (a)~bunch peak energy (blue circles) and FWHM energy spread (red squares), (b)~FWHM beam charge (blue circles) and FWHM normalized emittance (red squares), and (c)~FWHM transverse size (blue circles) and FWHM divergence (red squares). The fixed parameters are $n_u = 0.007n_c \simeq 1.2 \times 10^{19}$~cm$^{-3}$, $d = 0$ and $h = 10 \lambda_0 = 8~\mu$m. The laser pulse has propagated a distance of $\sim24\lambda_u$ in the underdense plasma region, which corresponds to a total propagation time of $t = 298\tau_0$ for $L_H = 6 \lambda_0$, $t = 303\tau_0$ for $L_H = 12 \lambda_0$, and $t = 310\tau_0$ for $L_H = 18 \lambda_0$.}
\label{fig:ParametricStudy_LH}
\end{figure*}

\begin{figure*}[bt]
    \centering
    \includegraphics[width=0.95\linewidth]{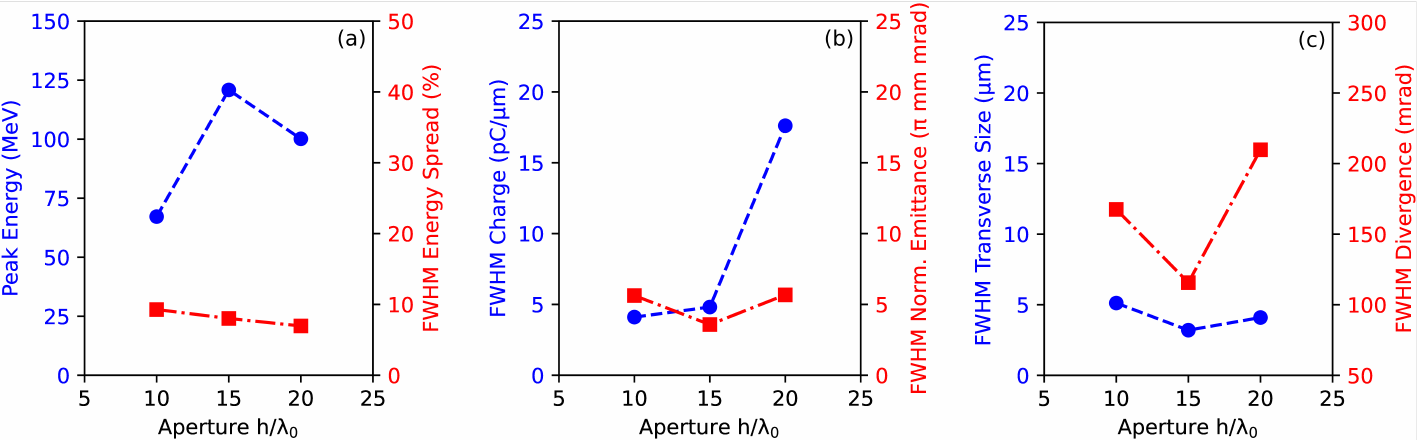}
    \caption{Parametric study as a function of the normalized overdense plasma aperture diameter $h / \lambda_0$ for electrons accelerated in the first wakefield cavity: (a)~bunch peak energy (blue circles) and FWHM energy spread (red squares), (b)~FWHM beam charge (blue circles) and FWHM normalized emittance (red squares), and (c)~FWHM transverse size (blue circles) and FWHM divergence (red squares). The fixed parameters are $n_u = 0.007n_c \simeq 1.2 \times 10^{19}$~cm$^{-3}$, $d = 0$ and $L_H = 12 \lambda_0 = 9.6~\mu$m. The laser pulse has propagated a distance of $\sim24\lambda_u$ in the underdense plasma region, which corresponds to a total propagation time of $t = 303\tau_0$ for all values of $h$.}
\label{fig:ParametricStudy_h}
\end{figure*}

A parametric study was subsequently conducted to evaluate the impact of key parameters on the properties of the accelerated electron beams: the underdense plasma density $n_u$, the gap distance $d$ between the overdense and underdense plasma, the overdense plasma length $L_H$, and the overdense plasma aperture diameter $h$. Figures~\ref{fig:ParametricStudy_nu}-\ref{fig:ParametricStudy_h} summarize the results for electrons accelerated in the first wakefield cavity (C1) after the laser pulse propagates a distance of $\sim 24 \lambda_0 / \sqrt{n_u/n_c}$ in the underdense plasma region, which corresponds to $t = 303\tau_0$ for the particular case $n_u = 0.007 n_c$ and $L_H = 12\lambda_0$, with $\omega_u = \sqrt{n_u e^2 / \varepsilon_0 m_e}$ the underdense plasma frequency and $\lambda_u = 2 \pi c / \omega_u = \lambda_0 \omega_0 / \omega_u = \lambda_0 / \sqrt{n_u/n_c}$ the underdense plasma wavelength. The high computational cost makes it prohibitive to carry out simulations for all the parameters until the electron beam reaches its maximum energy. However, it is important to note that the propagation distance considered ($\sim 24 \lambda_0 / \sqrt{n_u/n_c}$ in the underdense plasma) is long enough for the acceleration trends to be well established. This was verified in the case of some simulations running for longer times, as discussed in Section~\ref{Sec:OptimizedAcceleration}. Thus, the tendencies in beam energy, charge and quality analyzed in this section continue to be observed up to the maximum energy that can be achieve by the electron beam.

The underdense plasma density $n_u$ defines the underdense plasma wavelength $\lambda_u \propto n_u^{-1/2}$ and consequently determines the electron bunch capture location within the wakefield structure and energy transfer efficiency. Note that fixing the laser propagation distance in the underdense plasma at $\sim24\lambda_u$ and varying $n_u$ causes the propagation time to vary as well: $t = 457\tau_0$ for $n_u / n_c = 0.003$, $t = 303\tau_0$ for $n_u / n_c = 0.007$, and $t = 214\tau_0$ for $n_u / n_c = 0.015$.

The impact of altering $n_u$, while keeping the other simulation parameters unchanged, on the acceleration is presented in Fig.~\ref{fig:ParametricStudy_nu}. The FWHM charge increases with $n_u$, reaching $20.6$~pC/$\mu$m for $n_u / n_c = 0.015$. The peak energy also increases with $n_u$, from 21~MeV for $n_u / n_c = 0.003$ to 67~MeV for $n_u / n_c = 0.007$ and to 70~MeV for $n_u / n_c = 0.015$. Such result is not in accordance with the theoretical scaling law of Ref.~\cite{Lu2007}, which predicts that the maximum energy gain in a wakefield scales with $(2/3)(n_c/n_u)a_0$, indicating that lower plasma densities allow for higher final energies. However, this energy scaling can not be used on its own since Ref.~\cite{Lu2007} also estimates approximate optimal conditions for the bubble regime as $n_u/n_c = (\lambda_0 / \lambda_u)^2 \simeq a_0 (\lambda_0 / \pi W)^2 \approx 0.0063$ and $n_u/n_c = (\lambda_0 / \lambda_u)^2 \lesssim (\lambda_0 / 2 c \tau)^2 = 0.01$. In terms of beam quality, the proposed mechanism shows an optimal density around $n_u/n_c = 0.007$, which agrees with these estimations. For $n_u/n_c = 0.007$, the beam achieves low values of FWHM energy spread ($9\%$), emittance ($5.6~\pi~$mm~mrad), size ($5.1~\mu$m) and divergence ($168$~mrad). At lower and higher densities, the wakefield cavities do not form properly for effective acceleration. At $n_u = 0.003n_c$, despite the theoretical energy advantage, the wakefield structure is not completely formed for effective electron capture and acceleration, such that the beam presents a low quality with a FWHM energy spread of $45\%$, size of $22.4~\mu$m and divergence of $228$~mrad. At $n_u = 0.015n_c$, the shorter plasma wavelength reduces the acceleration cavity size and worsens the beam quality, which presents a FWHM energy spread of $24\%$ and divergence of $277$~mrad. {\it Thus $n_u / n_c = 0.007$ seems to be the better compromise for electron beam optimization and will be retained in the following.}

Varying the gap $d$ between the overdense and underdense plasma enables control over the relative position where target electrons are injected into the wakefield. This modifies the intensity of both longitudinal and transverse electric fields experienced by the injected electrons, directly affecting beam properties as shown in Fig.~\ref{fig:ParametricStudy_d}. Here once again the laser propagation distance in the underdense plasma is fixed at $\sim24\lambda_u$. But this time, the total propagation time varies due to the different values of the gap distance $d$: $t = 303\tau_0$ for $d = 0$, $t = 308\tau_0$ for $d = 5 \lambda_0$, and $t = 313\tau_0$ for $d = 10 \lambda_0$.

Increasing $d$ from $0$ to $10\lambda_0$ leads to a decrease in beam peak energy of approximately $14\%$, from $67$~MeV to $58$~MeV, and an increase in charge of $43\%$, from $4.1$~pC/$\mu$m to $5.9$~pC/$\mu$m, when considering the charge integrated over the FWHM energy range. Simultaneously, all the parameters related to the beam quality worsen. The FWHM energy spread increases from $9\%$ to $13\%$, the FWHM normalized emittance doubles from $5.6~\pi$~mm~mrad to $10.4~\pi$~mm~mrad, the FWHM transverse beam size increases from $5.1~\mu$m to $5.8~\mu$m (with a maximum of $6.4~\mu$m for $d = 5 \lambda_0$), and the FWHM beam divergence increases $18\%$, from $168$~mrad to $198$~mrad. {\it These results indicate that a smaller gap distance $d$ favors higher beam energy and better beam quality ({\it i.e.,} lower FWHM energy spread, emittance, beam size and divergence), while a larger $d$ enhances the beam charge at the expense of energy and beam quality.}

\begin{figure}[tb]
\centering
   \includegraphics[width=0.8\linewidth]{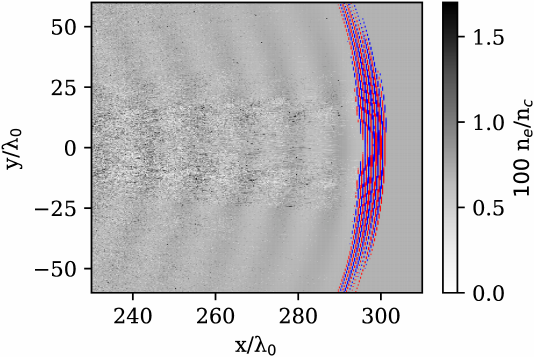}
    \caption{Laser-plasma interaction showing the electron density (grayscale) and laser pulse (red/blue) in $x$-$y$ space at $t=303\tau_0$ for $h = 5\lambda_0$, $n_u = 0.007n_c \simeq 1.2 \times 10^{19}$~cm$^{-3}$, $d = 0$ and $L_H = 12 \lambda_0 = 9.6~\mu$m.}
\label{fig:ElectronDensity_h5l0}
\end{figure}

Considering the overdense plasma alone, the energy gain of electrons extracted by the diffracted laser fields scales as $\gamma = 2 \eta a_0 \sqrt{k_0 L_H}$~\cite{Marini2023}, where $\eta$ is the simulated ratio between the diffracted field and the laser field maximum amplitudes. This scaling law describes only the laser-overdense plasma interaction and subsequent electron extraction and pre-acceleration, and it is consistent with the energy observed in the simulations. The difference in energy as a function of $L_H$ modifies the electron injection position into the wakefield and the amount of charge that can be effectively captured and accelerated. However, the injection energy of electrons coming from the overdense plasma is not a prediction of their final energy, since they will go through a second acceleration stage governed by the wakefield in the underdense region.

The impact of varying $L_H$ on the whole acceleration process is presented in Fig.~\ref{fig:ParametricStudy_LH}. Increasing $L_H$ from $6\lambda_0$ to $18\lambda_0$ results in a $19\%$ reduction in beam peak energy, from $73$~MeV to $60$~MeV, and a $71\%$ increase in the FWHM charge, from $3.0$~pC/$\mu$m to $5.1$~pC/$\mu$m. It is emphasized that the initial energy of electrons extracted from the overdense plasma (which scales with $\sqrt{L_H}$~\cite{Marini2023}) does not directly correlate with the final beam energy, as the injection position within the wakefield plays a more important role. On the other hand, charge extraction increases at larger $L_H$ values, which partially explains the reduction in final energy. The FWHM energy spread is $\sim15-16\%$ for $L_H = 6\lambda_0$ and $L_H = 18\lambda_0$, but reaches a minimum of $9\%$ for $L_H = 12\lambda_0$. Similarly, the FWHM normalized emittance is about $9-11~\pi~$mm~mrad for the lower and higher values of $L_H$, but it is minimized to $5.6~\pi~$mm~mrad at $L_H = 12\lambda_0$. Both FWHM beam transverse size and FWHM divergence do not vary between $L_H = 6\lambda_0$ and $L_H = 12\lambda_0$, but increase for $L_H = 18\lambda_0$, passing from $5.1~\mu$m to $6.4~\mu$m, and from $168$~mrad to $175$~mrad respectively. {\it These results suggest that an optimal target length exists around $L_H = 12\lambda_0$, where a balance between beam energy, charge, size and divergence is achieved, along with improved energy spread and emittance.}

The effect of the overdense plasma target aperture diameter $h$ was also investigated and is presented in Fig.~\ref{fig:ParametricStudy_h}. For $h$ smaller than $6\lambda_0$, a significant portion of the laser with beam waist $W = 9\lambda_0$ is blocked by the overdense plasma. The strong diffraction of the laser beam at the exit of the aperture prevents wakefield formation in the underdense plasma, as shown in Fig.~\ref{fig:ElectronDensity_h5l0} for $h = 5\lambda_0$, resulting in negligible electron acceleration. Conversely, when $h$ greatly exceeds the laser waist $W$, insufficient electrons are extracted from the overdense plasma edges for effective injection, as was observed for $h = 30\lambda_0$.

\setlength{\tabcolsep}{10.0pt}
\renewcommand{\arraystretch}{1.15}
 \begin{table*} [tb]
\caption{Beam parameters in the first wakefield cavity at maximum energy.}
  \centering
  \footnotesize
  \begin{tabular}{lcccc}
    \toprule
    \multirow{2}{*}{\textbf{Parameter}} & \textbf{$h = 10 \lambda_0$} & \textbf{$h = 15 \lambda_0$} & \multicolumn{2}{c}{\textbf{$h = 20 \lambda_0$}} \\
    [2pt] & \textbf{ } & \textbf{ } & \textbf{P1} & \textbf{P2} \\
    \midrule
    Time to reach Maximum Energy ($\tau_0$)             & 893   & 973   & \multicolumn{2}{c}{686}  \\
    Maximum Energy (MeV)                                & 180   & 278   & \multicolumn{2}{c}{224}  \\
    Peak Energy (MeV)                                   & 168   & 247   & 150   & 92    \\
    FWHM Energy Spread (MeV)                            & 8.8   & 10.9  & 9.5   & 14.1  \\
    FWHM Energy Spread ($\%$)                           & 5.2   & 4.4   & 6.3   & 15.3  \\
    FWHM Divergence (mrad)                              & 87    & 93    & 216   & 368   \\
    FWHM Transverse Size in $y$ direction ($\mu$m)      & 4.6   & 3.6   & 4.1   & 7.2   \\
    FWHM Normalized Emittance ($\pi$ mm mrad)           & 7.4   & 5.3   & 9.9   & 8.4   \\
    Integrated FWHM Charge (pC/$\mu$m)                  & 4.3   & 5.9   & 16.6  & 24.2  \\
    \hspace{10.0pt} From Overdense Plasma ($\%$)        & 100   & 93    & 88    & 6     \\
    \hspace{10.0pt} From Underdense Plasma ($\%$)       & 0     & 7     & 12    & 94    \\
    Estimated Total FWHM Charge (pC)                    & 54    & 107   & 396   & 356   \\
    Integrated Charge above 50~MeV (pC/$\mu$m)          & 8.0   & 19.9  & \multicolumn{2}{c}{105.6}  \\
    \hspace{10.0pt} From Overdense Plasma ($\%$)        & 100   & 76    & \multicolumn{2}{c}{34}     \\
    \hspace{10.0pt} From Underdense Plasma ($\%$)       & 0     & 24    & \multicolumn{2}{c}{66}     \\
    Estimated Total Charge above 50~MeV (pC)            & 101   & 328   & \multicolumn{2}{c}{1883}   \\
    Integrated Charge above 100~MeV (pC/$\mu$m)         & 8.0   & 19.9  & \multicolumn{2}{c}{62.7}   \\
    \hspace{10.0pt} From Overdense Plasma ($\%$)        & 100   & 76    & \multicolumn{2}{c}{55}     \\
    \hspace{10.0pt} From Underdense Plasma ($\%$)       & 0     & 24    & \multicolumn{2}{c}{45}     \\
    Estimated Total Charge above 100~MeV (pC)           & 101   & 328   & \multicolumn{2}{c}{1265}   \\
    \bottomrule
  \end{tabular}
  \label{tab:beam_params_max_ener}
\end{table*}

For moderate values of $h$, the aperture radius artificially imposes a waist on the laser pulse, changing the conditions for wakefield formation, particularly in the bubble regime. It allows a balance between effective electron extraction from the overdense plasma and suitable wakefield formation for reasonable energy gain, beam charge and quality. Figure~\ref{fig:ParametricStudy_h} shows that a maximum peak energy of 121~MeV is achieved for $h = 15\lambda_0 \sim 1.7 W$, whereas the charge greatly increases for $h = 20\lambda_0 \sim 2.2 W$, reaching $17.6$~pC/$\mu$m in the FWHM energy range. The FWHM energy spread decreases with $h$, from $9\%$ for $h = 10\lambda_0$ to $7\%$ for $h = 20\lambda_0$. The other quality parameters present a minimum at $h = 15\lambda_0$, where the FWHM beam emittance is $3.6~\pi~$mm~mrad, the FWHM size is $3.2~\mu$m and the FWHM divergence is $116$~mrad. {\it These results show that by varying the overdense plasma aperture diameter $h$ it is possible to obtain the highest values of peak energy and FWHM charge while keeping a reasonable beam quality.}

For the simulations presented in this section, the laser is focused at the center of the underdense region, {\it i.e.} $x_f=L_H+d+L_u/2$. Other values of the laser focal position $x_f$ were also investigated: $x_f = 0$ (at the entrance of the cylindrical aperture) and $x_f = L_H$ (at the exit of the cylindrical aperture). However, the final results did not present significant differences with varying values of $x_f$. This is due to the cylindrical aperture, which artificially imposes a waist on the laser pulse propagating in the underdense plasma, and to the existence of a second acceleration stage (much longer than the first one) responsible for the final electron beam properties.


\section{Optimized Laser-Plasma Electron Acceleration}
\label{Sec:OptimizedAcceleration}

\begin{figure*}[tb]
    \centering
    \includegraphics[width=0.295\linewidth]{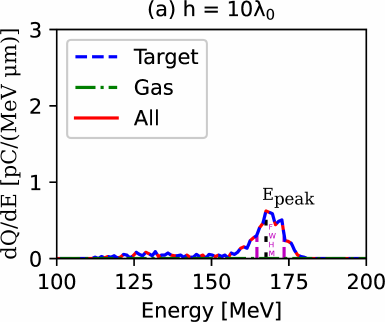}
    \includegraphics[width=0.295\linewidth]{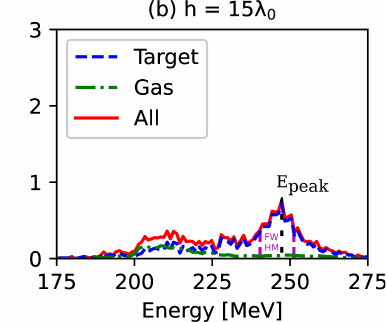}
    \includegraphics[width=0.35\linewidth]{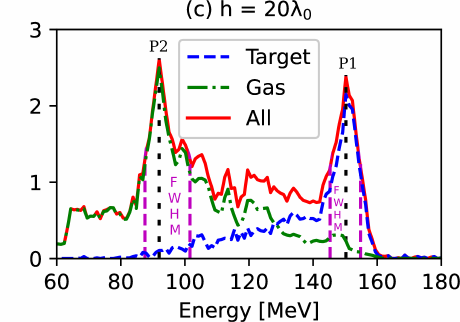}
    \caption{Electron energy spectra in the first wakefield cavity at maximum peak energy for (a) $h = 10\lambda_0$, $t = 893\tau_0$, (b) $h = 15\lambda_0$, $t = 973\tau_0$, and (c) $h = 20\lambda_0$, $t = 686\tau_0$. The dashed blue curve represents electrons from the overdense plasma (target), while the dash-dotted green curve shows electrons from the underdense plasma (gas). The solid red curve indicates the total electron population (all). The vertical dotted black line marks the peak energy, and the vertical dashed violet lines indicate the FWHM energy range.}
    \label{fig:ChargeDistributionMaxEnergy}
\end{figure*}

This section considers the parameters for which the electron beam presents a higher quality ({\it i.e.,} $n_u = 0.007n_c \simeq 1.2 \times 10^{19}$~cm$^{-3}$, $d = 0$, $L_H = 12 \lambda_0 = 9.6~\mu$m and $x_f = L_H + d + L_u/2 = 165 \lambda_0 = 132~\mu$m). As shown in Section~\ref{Sec:ParametricStudy}, the overdense plasma aperture diameter $h$ exerts the greatest influence on the maximum peak energy and maximum charge that can be achieved by the electron beam, and the trends observed in beam energy, charge and quality persist for longer time scales. For this reason, the role of parameter $h$ in the acceleration process is further investigated in this section at the moment the beam achieves the highest peak energy. Table~\ref{tab:beam_params_max_ener} summarizes the results, while Fig.~\ref{fig:ChargeDistributionMaxEnergy} displays the respective energy spectra.

For $h = 10 \lambda_0 = 8~\mu$m, the first wakefield cavity only captures electrons coming from the overdense plasma. At $t = 893 \tau_0$, the electron beam reaches its maximum peak energy and its quality is significantly improved. Compared to earlier instants of time, as in Table~\ref{tab:beam_params_303t0}, the beam achieves notably higher maximum ($180~$MeV) and peak energies ($168~$MeV) and narrower relative FWHM energy spread ($5.2\%$). Improved beam collimation is observed, indicated by a reduced FWHM divergence ($87~$mrad) and compact FWHM transverse profile ($4.6~\mu$m). These enhancements occur without compromising the beam charge ($4.3~$pC$/\mu$m) integrated over the FWHM energy range, thereby improving the overall beam quality. Considering all the electrons accelerated to energies above 100~MeV, the charge is almost doubled, reaching $8.0~$pC$/\mu$m.

The beam maximum and peak energies greatly increase for $h = 15 \lambda_0 = 12~\mu$m and $t = 973 \tau_0$, reaching $278~$MeV and $247~$MeV respectively. The amount of charge accelerated above 100~MeV is considerably higher ($19.9~$pC$/\mu$m) since electrons coming from both the overdense ($\sim 76\%$) and underdense ($\sim 24\%$) plasma are now accelerated. Considering the FWHM charge, the great majority of electrons come from the overdense plasma ($\sim 93\%$) and only $\sim 7\%$ from the underdense plasma, totalizing $5.9~$pC$/\mu$m. Despite the higher FWHM charge, the beam quality is enhanced, as can be seen by the lower relative FWHM energy spread ($4.4\%$), FWHM transverse size ($3.6~\mu$m) and FWHM emittance ($5.3~\pi~$mm~mrad).

The electron beam reaches its maximum peak energy faster, at $t = 686 \tau_0$, for $h = 20 \lambda_0 = 16~\mu$m. The first cavity also accelerates electrons from the overdense and the underdense plasma. However, it presents two peaks (labeled P1 and P2 in Table~\ref{tab:beam_params_max_ener} and Fig.~\ref{fig:ChargeDistributionMaxEnergy}(c)) in its energy spectrum. The peak energy P1 of $150~$MeV is lower in this case when compared to smaller values of $h$. Even though the FWHM transverse size ($4.1~\mu$m) is comparable, the general beam quality is worse, presenting $6.3\%$ of relative FWHM energy spread, $216$~mrad of FWHM divergence and $9.9~\pi~$mm~mrad of FWHM normalized emittance. This deterioration in beam quality is due to the much higher FWHM charge accelerated in this peak: $16.6~$pC$/\mu$m, corresponding to an estimated $396~$pC of FWHM charge in 3D.

While in the first peak (P1) $\sim 88\%$ of the accelerated electrons come from the overdense plasma, in the second energy peak (P2) only $\sim 6\%$ of the FWHM charge originates from the overdense plasma. Most of the charge ($\sim 94\%$) in the second peak comes from the underdense plasma, and it contains a comparable FWHM charge than P1: $24.2~$pC$/\mu$m, which corresponds to $356~$pC in 3D. On the other hand, the peak energy and beam quality are reduced. The second peak P2 presents a peak energy of $92~$MeV, with $15.3\%$ of FWHM energy spread, $368$~mrad of FWHM divergence and $7.2~\mu$m of FWHM size.

Considering both energy peaks P1 and P2, the total charge accelerated to energies above 50~MeV and 100~MeV increases considerably, reaching $105.6~$pC$/\mu$m and $62.7~$pC$/\mu$m respectively, with an estimated $1883~$pC and $1265~$pC in 3D. These values are much higher than the ones for $h = 10 \lambda_0$ and $h = 15 \lambda_0$: $8.0~$pC$/\mu$m ($101~$pC in 3D) and $19.9~$pC$/\mu$m ($328~$pC in 3D) respectively.

As an order-of-magnitude benchmark, we consider the 3D phenomenological nonlinear theory of Ref.~\cite{Lu2007} obtained in a pure wakefield regime. In that work, the theoretical electron beam energy is given by $(2/3) (n_c/n_u) a_0 m_e c^2 \approx 243$~MeV for $n_u=0.007n_c$ and $a_0=5$, which agrees very well with the peak energy of $247$~MeV obtained by simulations for $h = 15 \lambda_0$. In the case of $h = 10 \lambda_0 < 2W = 18 \lambda_0$, a considerable portion of the laser is blocked by the overdense target and does not contribute to the wakefield acceleration in the underdense plasma, which explains the lower peak energy. For $h = 20 \lambda_0 \approx 2W = 18 \lambda_0$, the maximum energy agrees with the theoretical estimate, but the peak energy is reduced due to the much higher amount of charge accelerated. It is also important to point out that the lower energies observed in the parametric study of Section~\ref{Sec:ParametricStudy} reflect the shorter propagation distance ($\sim 24\lambda_u$ in the underdense plasma), {\it i.e.} before dephasing and thus before the electron beam can reach its maximum energy.

Reference~\cite{Lu2007} estimates the number of electrons accelerated in the first wakefield cavity as
\begin{equation}
    N_{e,C1} \simeq 2.5 \times 10^9 \left(\frac{\lambda_0~[\mu\mathrm{m}]}{0.8}\right) \sqrt{\frac{P~[\mathrm{TW}]}{100}} \, ,
    \label{eq:N_e,C1}
\end{equation}
where $P = \pi I_0 W^2 / 2$ for a Gaussian beam. For the scheme proposed in this paper, part of the laser can be reflected back by the overdense plasma as shown in Fig.~\ref{fig:ElectronDensityEvolution}(a). Therefore, the power should be rewritten as $P = \pi I_0 W_\mathrm{eff}^2 / 2$, with $W_\mathrm{eff} = h/2$ if $h/2 < W$ and $W_\mathrm{eff} = W$ if $h/2 \ge W$.

Following Eq.~(\ref{eq:N_e,C1}), the electron charge $Q_{e,C1} = e N_{e,C1}$ in the first wakefield cavity is estimated as $Q_{e,C1} \approx 146$~pC for $h = 10 \lambda_0$, $Q_{e,C1} \approx 220$~pC for $h = 15 \lambda_0$ and $Q_{e,C1} \approx 263$~pC for $h = 20 \lambda_0$. In the case $h = 10 \lambda_0$, the charge ($146$~pC) estimated by the theoretical scaling law of Ref.~\cite{Lu2007} is higher than the one ($101$~pC) projected in the proposed acceleration scheme due to the fact that $h \sim W$ and, consequently, a great portion of the laser energy is blocked by the overdense target and do not interact with the underdense plasma for wakefield formation and acceleration. On the other hand, for $h = 15 \lambda_0 \lesssim 2W$, the total charge ($328$~pC) accelerated by the proposed scheme is higher than the theoretical estimate ($220$~pC). And for $h = 20 \lambda_0 \gtrsim 2W$, the theoretical charge ($263$~pC) is much lower than the total charge ($1883$~pC) projected from the simulations. This happens because, for $h \sim 2W$, only a small fraction of the laser energy is blocked by the overdense plasma and electrons coming from both the overdense and the underdense plasma are injected in the laser-driven wakefield for acceleration, in contrast with the theoretical estimates which consider a pure laser-underdense plasma interaction.

It is important to note that the scalings in Ref.~\cite{Lu2007} are presented here as a benchmark rather than an exact prediction because they were derived for an ideal three-dimensional nonlinear bubble in an underdense plasma, whereas the simulations in this paper are two-dimensional and the laser is modified by the overdense target aperture. A single closed-form model is not expected to reproduce the complete two-stage acceleration process since electron extraction from the overdense plasma and the subsequent injection and acceleration of electrons from both the overdense and the underdense plasma into the wakefield are intrinsically coupled.

{\it The results in this section indicate that the proposed injection and acceleration scheme is effective at producing high-charge energetic electron beams over sub-millimeter distances. Furthermore, the overdense plasma aperture diameter $h$ can be used as a control parameter to tune the beam properties, delivering higher energies around $h = 15 \lambda_0$ and higher amounts of charge around $h = 20 \lambda_0$.}


\section{Conclusions and Discussion}
\label{Sec:Conclusion}

Different mechanisms exist for electron acceleration in plasmas. In the case of overdense plasmas, $\sim$nC of charge can be achieved with a few MeV, but the beam presents a poor quality. Underdense plasmas, on the other hand, are capable of accelerating electrons up to the GeV level with high quality, but the charge is low (a few pC). This paper presents an alternative electron acceleration scheme combining diffracted wave excitation from laser–overdense plasma interaction with wakefield generation in an underdense plasma. From 2D3V particle-in-cell simulations carried out with the \Smilei code, it was demonstrated that electrons extracted from the overdense plasma can be effectively injected into the underdense wakefield and accelerated to high energies, forming electron bunches with peak energies of $\sim150-250$~MeV, estimated FWHM charge of $\sim50-400$~pC, and $\sim100-1900$~pC of charge with energies above $50$~MeV.

Through a parametric study, the impact of key parameters on beam properties including peak energy, FWHM energy spread, charge, emittance, transverse size and divergence was analyzed. It was shown that the underdense plasma density plays an essential role in controlling the beam quality. A small gap distance between the overdense and underdense plasma produces beams with higher quality and higher energy, whereas increasing the gap distance increases the beam charge but reduces its energy and quality. Intermediate values of overdense plasma length allow for a better balance between beam energy, charge and quality.

The most important parameter analyzed for the proposed scheme is the overdense plasma aperture diameter $h$. For $h = 10\lambda_0$, only electrons coming from the overdense plasma are accelerated in the first wakefield cavity, reaching, at laser intensity $I_0\lambda_0^2 \simeq 3.4 \times 10^{19}$~W$\mu$m$^2$/cm$^2$, $168$~MeV of peak energy with an estimated charge of $54$~pC in the FWHM energy range and $101$~pC with energies above $100$~MeV. For higher values of $h$, electrons form both the overdense and the underdense plasma are accelerated in the first wakefield cavity. An aperture diameter of $h = 15 \lambda_0$ produces the most energetic electron beams, with $247$~MeV at peak, $107$~pC of FWHM charge and $328$~pC above $100$~MeV. Increasing $h$ to $20 \lambda_0$ allows the acceleration of higher amounts of charge, reaching up to $1265$~pC for energies above $100$~MeV, but produces an energy spectrum with two energy peaks: one at $150$~MeV with $396$~pC of FWHM charge, and another one at $92$~MeV and $356$~pC of FWHM charge.

Understanding and optimizing the overdense and underdense plasma parameters enables the generation of electron beams with improved quality while maintaining substantial charge and high energy, offering an alternative approach for advanced electron sources. From the results presented in this paper, it appears that the parameters to be optimized depend on the choice between obtaining maximum energy or maximum charge.
 
Further optimization of this acceleration scheme is still possible and is currently in progress. Exploring a wider parameter space could enable the production of beams with improved quality, higher energy and/or higher charge. In particular, adding a transverse profile in the underdense plasma density would enable the laser to propagate as in a waveguide, thus allowing the study of the proposed mechanism while avoiding laser diffraction and depletion.

Different effects would require going beyond the present 2D3V simulations, such as the azimuthal variations expected from the linear polarization of the laser, and out-of-plane quasi-static magnetic components, which can exist in a 2D model but their complete azimuthal topology and its influence on beam collimation and injection can only be represented in three dimensions. The study of these effects, and more precise electron beam parameters values, would involve full 3D simulations. Such simulations are out of the scope of this paper due to the prohibitive computational resources required by the need for high spatial and temporal resolutions, high number of macro-particles per cell and large simulation box for accurate results. Nonetheless, the performed 2D3V simulations are able to describe well enough the essential physical mechanisms and qualitative trends for electron extraction from the overdense plasma, and their injection and further acceleration in the wakefield together with electrons originating from the underdense plasma. The same is observed in other 2D/3D studies of diffracted waves~\cite{Marini2023} and wakefield acceleration~\cite{Vieira2008, Zeng2014, Fedeli2022}, with full 3D simulations mainly differing from the 2D ones in the quantitative values of beam properties.

It is also worth to mention that other laser polarizations could be used, but they would produce different effects. A circular polarization, for example, would change the electron heating and extraction processes and could generate a longitudinal magnetic field through the inverse Faraday effect~\cite{Steiger1972, Shvets2002}. Whereas a driver carrying orbital angular momentum (a Laguerre-Gaussian laser pulse) could create donut shaped wakefield cavities and produce hollow or helical electron bunches~\cite{Vieira2014}.

Finally, it is important to note that it is possible to carry out experiments with the proposed acceleration mechanism in current facilities delivering short ($\sim$fs) laser pulses with intensities $\ge 10^{19}$~W/cm$^2$. The overdense plasma is a solid target with a micrometer-scale cylindrical aperture of radius $h/2 \sim 4$–$8~\mu$m, which can be produced by standard micro-fabrication processes. The cylindrical aperture sets the transmitted laser waist and, thus, must be aligned with the laser axis. The acceleration scheme is robust regarding the formation of a pre-plasma in front of the solid target or inside its aperture, but it could alter the laser diffraction and electron extraction. Therefore, a high laser contrast or a plasma mirror is preferable. The underdense plasma region should be a quasi-uniform gas, produced by a gas nozzle or cell, that extends behind the solid target.


\acknowledgments

This project has received funding from the European Union’s Horizon 2020 research and innovation programme under the Marie Sk\l{}odowska-Curie grant agreement No 899987, and Plas@Par Grant No. ANR-11-IDEX-0004-02. This project was provided with computing HPC and storage resources by GENCI at TGCC thanks to the grants 2013/2015/2017-0507678 on the supercomputer Joliot Curie's SKL and ROME partitions. Simulations were performed with the open-source particle-in-cell (PIC) code \Smilei. The authors are grateful to the TIPS-LULI team for fruitful discussions and to the \Smilei dev-team for technical support.


\section{Author Declarations}
\label{Sec:Author_Declarations}

\subsection{Conflict of Interest}
\label{Sec:Conflict_Interest}

The authors have no conflicts to disclose.


\section{Data Availability}
\label{Sec:Data_Availability}

The data that supports the findings of this study are available from the corresponding author upon reasonable request.

\bibliographystyle{apsrev4-2}
\bibliography{biblio}

\end{document}